\documentclass[amsmath,amssymb,showpacs,twocolumn,superscriptaddress,pr]{revtex4-1}
\usepackage{graphicx,bm,color,subfigure}
\usepackage{amsmath}
\usepackage{hyperref}
\newcommand{\bs}{\boldsymbol}

\begin{document}

\title{Singlet $s^\pm$-wave pairing in quasi-one-dimensional ACr$_3$As$_3$ (A=K, Rb, Cs) superconductors}

\author{Li-Da Zhang}
\thanks{These two authors contributed equally to this work.}
\affiliation{School of Physics, Beijing Institute of Technology, Beijing 100081, China}

\author{Xiaoming Zhang}
\thanks{These two authors contributed equally to this work.}
\affiliation{Institute for Advanced Study, Tsinghua University, Beijing 100084, China}

\author{Juan-Juan Hao}
\affiliation{School of Physics, Beijing Institute of Technology, Beijing 100081, China}

\author{Wen Huang}
\affiliation{Institute for Advanced Study, Tsinghua University, Beijing 100084, China}

\author{Fan Yang}
\email{yangfan\_blg@bit.edu.cn}
\affiliation{School of Physics, Beijing Institute of Technology, Beijing 100081, China}

\begin{abstract}
The recent discovery of quasi-one-dimensional Chromium-based superconductivity has generated much excitement. We study in this work the superconducting instabilities of a representative compound, the newly synthesized KCr$_3$As$_3$ superconductor. Based on inputs from density functional theory calculations, we first construct an effective multi-orbital tight-binding Hamiltonian to model its low-energy band structure. We then employ standard random-phase approximation calculations to investigate the superconducting instabilities of the resultant multi-orbital Hubbard model. We find various pairing symmetries in the phase diagram in different interaction parameter regimes, including the triplet $f$-wave, $p_z$-wave and singlet $s^\pm$-wave pairings. We argue that the singlet $s^\pm$-wave pairing, which emerges at intermediate interaction strength, may be realized in this material. This singlet pairing is driven by spin-density wave fluctuations enhanced by Fermi-surface nesting. We point out that phase-sensitive measurement can distinguish the $s$-wave pairing in KCr$_3$As$_3$ from the $p_z$-wave previously proposed for a related compound K$_2$Cr$_3$As$_3$. The $s^{\pm}$-wave pairing in KCr$_3$As$_3$ shall also exhibit a subgap spin resonance mode near the nesting vector, which can be tested by inelastic neutron scattering measurements.  Another intriguing property of the $s^\pm$-pairing is that it can induce time-reversal invariant topological superconductivity in a semiconductor wire with large Rashba spin-orbit coupling via proximity effect. Our study shall be of general relevance to all superconductors in the family of ACr$_3$As$_3$ (A=K, Rb, Cs).
\end{abstract}



\maketitle


\section{Introduction}
There has been a recent surge in the discovery of quasi-1D Cr- and Mo-based superconductors at ambient pressures \cite{K2Cr3As3,Rb2Cr3As3,Cs2Cr3As3,Mu:17,Liu:17,Mu:18a,Mu:18,Zhao:18}. The crystals of these compounds consist of alkali metal atom separated [(Cr$_3$(Mo)$_3$As$_3$)$^{2-}$]$_{\infty}$ double-walled subnanotubes. Their low-energy degrees of freedom are generally dominated by the Cr(Mo) 3d(4d) orbitals, with stronger electron correlations generally expected for Cr. In the present study, we focus on the Chromium family. The 3d orbitals form multiple bands, some with strong 1D character and some more 3D-like \cite{Jiang:15}, depending on the microscopic details. The electronic structure with multibands of 3d-electrons resembles the situation in iron-based superconductors. However, the distinct dimensionality foretells markedly different physical properties, including the superconductivity (SC), with some discussions of Cooper pairing born out of a Tomonaga-Luttinger liquid normal state \cite{Miao:16}.

These quasi-1D Cr-based compounds can be classified into two families with 133 and 233 chemical compositions, respectively:  ACr$_3$As$_3$ and A$_2$Cr$_3$As$_3$, where A=K, Rb, Cs stands for the alkaline elements. Following the discovery of SC in the 233 family\cite{K2Cr3As3,Rb2Cr3As3,Cs2Cr3As3}, these systems have received considerable interest. Several theoretical proposals, on the basis of effective multi-orbital Hubbard models and within the framework of either weak-coupling or strong-coupling descriptions, have provided support for spin-triplet $p_z$ or $f$-wave pairings \cite{Zhou:17,Wu:15,Zhang:16,Miao:16}. However, on the experimental fronts, multiple measurements have revealed conflicting signatures\cite{Pang:15,Balakirev:15,Yang:15,Zhi:15,Adroja:15,Adroja:17,Taddei:17}, hence no consensus has been achieved regarding its pairing symmetry. A serious drawback of the 233 family is that it is chemically unstable at ambient environment\cite{K2Cr3As3,Rb2Cr3As3,Cs2Cr3As3}, which brings difficulty to experimental measurements. This drawback, however, is absent in the newly synthesized 133 family\cite{Mu:17,Liu:17}, which are stable at ambient environment and thus allow for more thorough experimental investigation of their properties. Furthermore, while the 233 family is noncentrosymmetric, the 133 family is centrosymmetric. Thus far, there has been no report about the nature of the pairing in the latter compounds.

In this paper, we construct effective models to study the pairing symmetry of a representative Cr-based 133 superconductor, KCr$_3$As$_3$. Based on the electronic structure obtained from first principle calculations, we build an effective tight-binding (TB) model with Wannier orbitals of $d$-wave symmetries. Adopting general Hubbard-like interactions between the multiple Wannier d-orbitals, we then employ the random phase approximation (RPA) approach to evaluate the effective interactions in the Cooper channel, from which we obtain the pairing symmetry. This approach is a descendant of the celebrated Kohn-Luttinger mechanism \cite{Kohn:65}, and is generally considered valid in the weak-coupling limit within which SC is the only instability present in the system \cite{Raghu:10}.

\begin{figure*}
\includegraphics[width=7in]{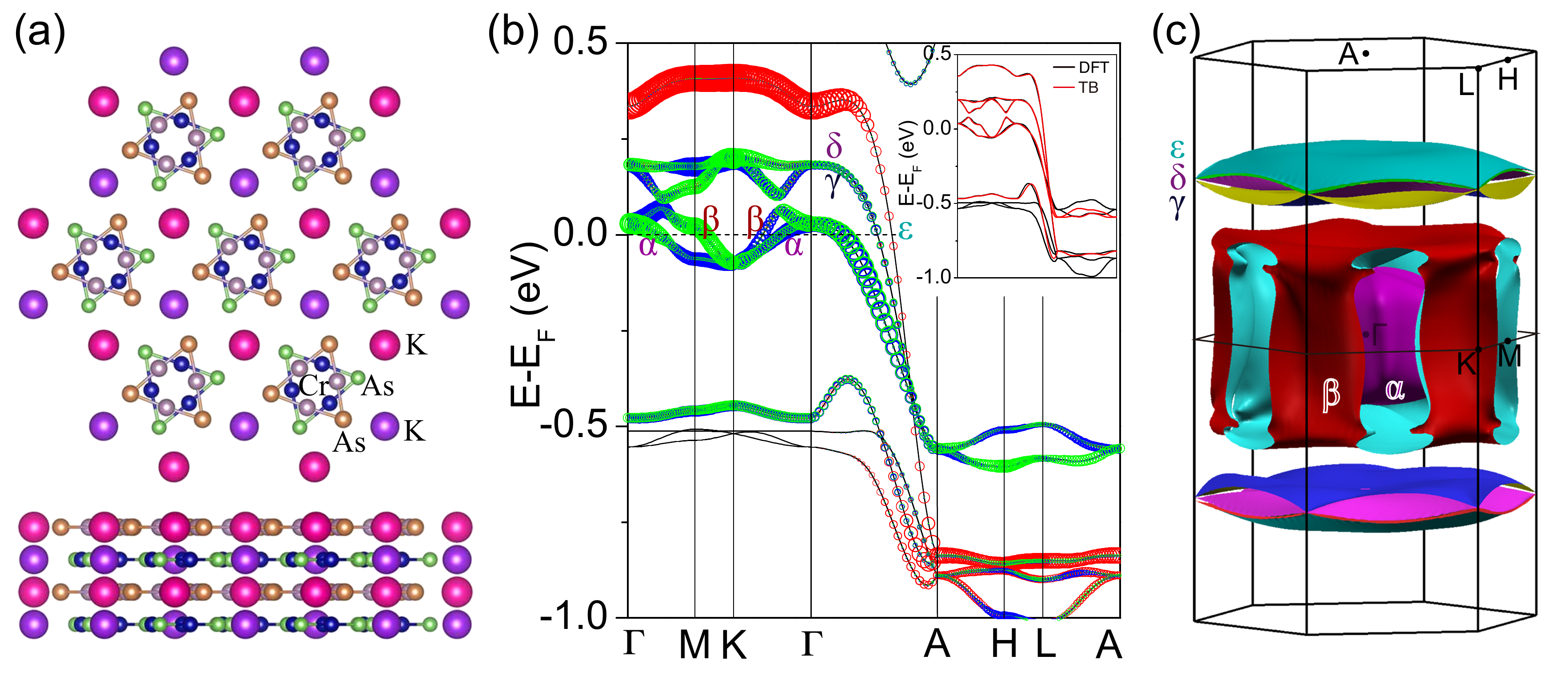}
\caption{(color online).(a) The top and side view for the crystal structure of KCr$_3$As$_3$. (b) Band structure of KCr$_3$As$_3$ along the high symmetry lines, with the red/blue/green circles being drown proportional to the weight of Cr-3d$_{z^2}$/-3d$_{x^{2}-y^{2}}$/-3d$_{xy}$, respectively. The five bands crossing with the Fermi-level are marked as $\alpha$ - $\varepsilon$ respectively. The inset is the comparison between the band structures calculated by DFT and TB model. (c) The FSs of KCr$_3$As$_3$, with the high symmetric points and the band indices for the five FSs marked.}
\label{structure}
\end{figure*}

We obtain a phase diagram similar to that of K$_2$Cr$_3$As$_3$ \cite{Wu:15,Zhang:16}, except for the appearance of $s^\pm$-wave pairing in the intermediate Hubbard-$U$ regime. To be more concrete, the strong-$U$ limit is dominated by a spin-density-wave (SDW) phase; SC becomes the only instability at weaker $U$. In particular, at relatively strong Hund's rule coupling ($J_H/U>r_c\approx\frac{1}{3}$), an on-site $f$-wave pairing becomes more favorable; and at weaker Hund's rule coupling, $p_z$-wave pairing develops in the weak-$U$ limit and $s^\pm$-wave pairing dominates at intermediate $U$. However, as the $p_z$-wave pairing developed under the Kohn-Luttinger mechanism is too weak to be consistent with the experimental $T_c$, we argue here that only the $s^\pm$-wave pairing represents a reasonable candidate for this material. This singlet pairing is driven by SDW fluctuations enhanced by Fermi-suface- (FS-) nesting. We propose a phase-sensitive SQUID measurement to distinguish the $s$-wave pairing in KCr$_3$As$_3$ from the $p_z$-wave pairing in K$_2$Cr$_3$As$_3$.  The sign-changing $s^{\pm}$-wave state is expected to support an in-gap spin resonance mode near the nesting wavevector, which can be detected in inelastic neutron scattering experiments. Another remarkable property of the $s^\pm$-pairing is that it can induce, via proximity effect, time-reversal invariant (TRI) topological SC (TSC) in a semiconductor wire with large Rashba spin-orbit coupling (SOC).

The rest of this paper is organized as follows. In Sec. \ref{sec:DFT}, we provide our results from first-principle calculations based on density-functional theory (DFT) for the band structure of KCr$_3$As$_3$, after which we construct its effective TB model. In Sec. \ref{sec:phaseDiag}, we study the superconducting pairing symmetry of the system and present a phase diagram. In Sec. \ref{sec:sWave}, we focus on the properties of the obtained $s^\pm$-wave SC. Our results are summarized in Sec. \ref{sec:summary}.

\section{Band structure and the TB Model}
\label{sec:DFT}
\subsection{The DFT band structure}
\label{subsec:LDA}
The crystal structure of KCr$_3$As$_3$ and its sister compounds is shown in Fig.\ref{structure} (a), which belongs to the space group $P6_{3}/m$ (point group $C_{6h}$). Their basic building blocks are 1D chains of double-walled subnanotubes of [Cr$_3$As$_3$]$_\infty$. The Cr$_3$As$_3$ units stack in alternating fashion along the chain direction, such that each unit cell contains two sublayers, A and B. The chains are separated by the alkali cations.

The band structure of KCr$_3$As$_3$ was predicted by using DFT as implemented in the Vienna {\it ab initio} simulation pack (VASP) \cite{Kresse1,Kresse2}. We employed the generalized gradient approximation (GGA) in the form of Perdew-Burke-Ernzerhof (PBE) \cite{Perdew} to describe the exchange-correlation effects. The electron-ion interactions was simulated using projector-augmented-wave (PAW) potentials with the energy cutoff of 500 eV\cite{Kresse3}. Charge density calculations were performed on $8\times8\times12$ Monkhorst-Pack sampling in the Brillouin zone.

The band structure thus obtained is shown in Fig.\ref{structure} (b). One can clearly see that the five bands (marked as $\alpha, \beta, \gamma, \delta, \varepsilon$) crossing with the Fermi level are dominated by the 3d$_{z^2}$, the 3d$_{x^{2}-y^{2}}$ and the 3d$_{xy}$ orbitals of Cr atoms. Overall, these orbitals are mixed together but the relative contributions to the metallic bands are different. The 3d$_{x^{2}-y^{2}}$ and the 3d$_{xy}$ orbitals mix and mainly contribute to the two 3D FSs (which belong to the $\alpha$- and $\beta$- bands) and the two quasi-1D FSs (which belong to the $\delta$- and $\gamma$- bands), while the 3d$_{z^2}$ orbital is relative decoupled and forms one quasi-1D FS (which belong to the $\varepsilon$- band). The five FSs are plotted in Fig.\ref{structure} (c). The band structure and FS obtained here are consistent with those in a previous report\cite{Cao133}.

\subsection{The TB Model}
\label{subsec:TB}
To obtain an effective model, a sparse $4\times4\times6$ $\Gamma$-centered sampling is utilized for establishing Wannier orbitals of virtual atoms localized at the center of each Cr triangle via WANNIER90 package\cite{Mostofi}, with the initial guess of d$_{z^2}$ ($A'$ representation) and (d$_{x^{2}-y^{2}}$, d$_{xy}$) ($E'$ representation) orbitals for unitary transformations. The structure of KCr$_3$As$_3$ can then be simplified to be chains of virtual atoms arranged in parallel, with each chain containing A and B sublattices. The obtained TB Hamiltonian in the momentum space can be expressed as
\begin{align}
&H^{(0)}_{{\rm TB}}
=\sum_{\bm{k}mn\mu\nu\sigma}h^{mn}_{\mu\nu}(\bm{k})
c^{\dagger}_{m\mu\sigma}(\bm{k})c_{n\nu\sigma}(\bm{k})            \nonumber\\
=&\sum_{\bm{k}\mu\nu\sigma}
\left(\begin{array}{cc}
c^{\dagger}_{A\mu\sigma} & c^{\dagger}_{B\mu\sigma}
\end{array}\right)
\left(\begin{array}{cc}
h^{AA}_{\mu\nu} & h^{AB}_{\mu\nu} \\
h^{BA}_{\mu\nu} & h^{BB}_{\mu\nu}
\end{array}\right)
\left(\begin{array}{c}
c_{A\nu\sigma} \\ c_{B\nu\sigma}
\end{array}\right)
\end{align}
where $m,n=A,B$ label sublattices, $\sigma=\uparrow,\downarrow$ label spins and $\mu,\nu=1,2,3$ label orbitals $d_{z^2}$, $d_{xy}$ and $d_{x^2-y^2}$, respectively. Note that physically the orbital bases $d_{z^2}$, $d_{xy}$ and $d_{x^2-y^2}$ here
are not the maximally localized Wannier wave functions around a real single Cr atom, but the superposition of the three atomic wave functions around the virtual center. Such Wannier wave functions can be dubbed as ``molecular orbital"\cite{Zhou:17, Dai}, which are much more delocalized than conventional ``atomic-orbital" Wannier bases localizing around a single atom\cite{Zhou:17, Dai, Wu:15, Zhang:16}.

The minimum TB model which can capture all the qualitative aspects of the low energy part of the DFT band structure contains up to the next-nearest-neighbor hopping parameters in the $xy$-plane and the next-next-nearest-neighbor hopping parameters in the $z$-direction.  The point-group symmetry allowed $h^{mn}_{\mu\nu}(\bm{k})$ thus obtained can be expressed as following,
\begin{align}
h^{AA}_{\mu\nu}&=
\sum^3_{n=0}(2-\delta_{n,0})\cos(2nz)
\Bigg\{t^{0,n}_{\mu\nu}                             \nonumber\\
&+\Big[t^{1,n}_{\mu\nu}e^{2iy}
+p^{1,n}_{\mu\nu}e^{-i(x+y)}
+d^{1,n}_{\mu\nu}e^{i(x-y)}                         \nonumber\\
&~~+t^{2,n}_{\mu\nu}e^{2ix}
+p^{2,n}_{\mu\nu}e^{i(3y-x)}
+d^{2,n}_{\mu\nu}e^{-i(3y+x)}                         \nonumber\\
&~~+(\text{terms with }
\mu\leftrightarrow\nu,~x\rightarrow -x,
~y\rightarrow -y)\Big]\Bigg\}
\end{align}
\begin{align}
h^{AB}_{\mu\nu}&=
\sum^3_{n=0}2\cos[(2n+1)z]
\Bigg\{\tilde{t}^{0,n}_{\mu\nu}                     \nonumber\\
&+\Big[t^{1A,n}_{\mu\nu}e^{2iy}
+p^{1A,n}_{\mu\nu}e^{-i(x+y)}
+d^{1A,n}_{\mu\nu}e^{i(x-y)}                         \nonumber\\
&~~+t^{2A,n}_{\mu\nu}e^{2ix}
+p^{2A,n}_{\mu\nu}e^{i(3y-x)}
+d^{2A,n}_{\mu\nu}e^{-i(3y+x)}                         \nonumber\\
&~~+(\text{terms with }
\mu\leftrightarrow\nu,~A\rightarrow B,                        \nonumber\\
&~~~~~~~x\rightarrow -x,~y\rightarrow -y)\Big]\Bigg\}
\end{align}
\begin{align}
h^{BB}_{\mu\nu}=h^{AA*}_{\mu\nu},~~~~~~~~~~
h^{BA}_{\mu\nu}=h^{AB*}_{\mu\nu}
\end{align}
\begin{align}
p_{\mu\nu}=&\sum_{\mu^{\prime}\nu^{\prime}}U_{\mu\mu^{\prime}}t_{\mu^{\prime}\nu^{\prime}}U^{\dagger}_{\nu^{\prime}\nu}  \nonumber\\
d_{\mu\nu}=&\sum_{\mu^{\prime}\nu^{\prime}}U^{\dagger}_{\mu\mu^{\prime}}t_{\mu^{\prime}\nu^{\prime}}U_{\nu^{\prime}\nu}
\end{align}
and
\begin{align}
U=\left(\begin{array}{ccc}
1 & 0 & 0 \\
0 & -1/2 & \sqrt{3}/2 \\
0 & -\sqrt{3}/2 & -1/2 \\
\end{array}\right)
\end{align}
Here $x=\frac{\sqrt{3}}{2}{k_xa_0}$, $y=\frac{1}{2}k_ya_0$ and $z=\frac{1}{2}k_zc_0$. The values of the above parameters are listed in the Appendix.

The TB model actually adopted in our following calculations is a more rigourous one expressed as,
\begin{align}
H_{{\rm TB}}
=\sum_{\bm{k}\mu\nu\sigma}h_{\mu\nu}(\bm{k})
c^{\dagger}_{\mu\sigma}(\bm{k})c_{\nu\sigma}(\bm{k}),
\end{align}
with $\mu,\nu=1,\cdots, 6$ indicating the orbital-sublattice indices. The elements of the $h(\bm{k})$ matrix is,
\begin{align}
h_{\mu\nu}(\bm{k})=\sum_{r_1,r_2,r_3}t^{r_1,r_2,r_3}_{\mu\nu}e^{i\bm{k}\cdot(r_1\bm{a_1}+r_2\bm{a_2}+r_3\bm{a_3})}.\label{TB_parameters}
\end{align}
Here $\bm{a_1}=(\frac{\sqrt{3}}{2}a_0,-\frac{1}{2}a_0,0)$, $\bm{a_2}=(0,a_0,0)$ and $\bm{a_3}=(0,0,c_0)$. The data of $t^{r_1,r_2,r_3}_{\mu\nu}$ for $r_{1,2}\in[-2,2]$, $r_3\in[-4,4]$ and $\mu,\nu\in[1,6]$ is provided in the {\color{red}Supplementary Material \cite{SuppMat}}. The band structure thus obtained shown in the inset of Fig.~\ref{structure}(b) (red) is well consistent with that of the DFT (black) at low energy near the Fermi level.

\section{Pairing symmetries and the phase-diagram}
\label{sec:phaseDiag}
We adopt the following Hamiltonian in our study:
\begin{align}\label{model}
H=&H_{\text{TB}}+H_{int}\nonumber\\
H_{int}=&U\sum_{i\mu}n_{i\mu\uparrow}n_{i\mu\downarrow}+
V\sum_{i,\mu<\nu}n_{i\mu}n_{i\nu}+J_{H}\sum_{i,\mu<\nu}                   \nonumber\\
&\Big[\sum_{\sigma\sigma^{\prime}}c^{+}_{i\mu\sigma}c^{+}_{i\nu\sigma^{\prime}}
c_{i\mu\sigma^{\prime}}c_{i\nu\sigma}+(c^{+}_{i\mu\uparrow}c^{+}_{i\mu\downarrow}
c_{i\nu\downarrow}c_{i\nu\uparrow}+h.c.)\Big]
\end{align}
Here, the interaction parameters $U$, $V$, and $J_H$ denote the intra-orbital, inter-orbital Hubbard repulsion, and the Hund's rule coupling (as well as the pair hopping) respectively, which satisfy the relation $U=V+2J_H$. We tune $U$ and $J_H/U$ to obtain a phase diagram of the pairing instabilities and density wave orders.

Note that as the ``molecular-orbital" Wannier bases adopted here are much more delocalized than conventional local ``atomic-orbital" Wannier bases, the effective interaction parameters $U$, $V$, and $J_H$ defined here are much weaker than those defined for conventional local Wannier bases. While the former group of parameters are not easy to be estimated directly through first principle calculations, they are related to the latter group by a complicated formula under coarse approximations\cite{Dai}. Further more, although the latter group of parameters can be roughly estimated via such first principle approaches as the constrained DFT or the constrained RPA, the obtained values are strongly approach-dependent. Therefore, in our present study, we simply set these interaction parameters as tuning parameters to be determined by experimental results.

\subsection{The RPA approach}
\label{sec:RPA}
Following the standard multi-orbital RPA approach \cite{RPA1,RPA2,RPA3,Kuroki,Scalapino1,Scalapino2,Liu2013,Wu2014,Ma2014,Zhang2015,Liu2018}, we first define the following bare susceptibility in the normal state for the non-interacting case:
\begin{align}\label{chi0}
\chi^{(0)}_{pqst}(\bm{k},\tau)\equiv
&\frac{1}{N}\sum_{\bm{k}_1\bm{k}_2}\left\langle
T_{\tau}c_{p}^{\dagger}(\bm{k}_1,\tau)
c_{q}(\bm{k}_1+\bm{k},\tau)\right.                      \nonumber\\
&\left.\times c_{s}^{\dagger}(\bm{k}_2+\bm{k},0)
c_{t}(\bm{k}_2,0)\right\rangle_0,
\end{align}
Here $\langle\cdots\rangle_0$ denotes the thermal average for the noninteracting system, $T_{\tau}$ denotes the imaginary time-ordered product, and $p,q,s,t=1,\cdots,6$ are the orbital-sublattice indices. Fourier transformed to the imaginary frequency space, the bare susceptibility can be expressed by the following explicit formulism:
\begin{align}\label{chi0e}
\chi^{(0)}_{pqst}(\bm{k},i\omega_n)
=&\frac{1}{N}\sum_{\bm{k}'\alpha\beta}
\xi^{\alpha}_{t}(\bm{k}')
\xi^{\alpha*}_{p}(\bm{k}')
\xi^{\beta}_{q}(\bm{k}'+\bm{k})                         \nonumber\\
&\times\xi^{\beta*}_{s}(\bm{k}'+\bm{k})
\frac{n_F(\varepsilon^{\beta}_{\bm{k}'+\bm{k}})
-n_F(\varepsilon^{\alpha}_{\bm{k}'})}
{i\omega_n+\varepsilon^{\alpha}_{\bm{k}'}
-\varepsilon^{\beta}_{\bm{k}'+\bm{k}}}.
\end{align}
where $\alpha,\beta=1,\cdots,6$ are band indices, $\varepsilon^{\alpha}_{\bm{k}}$ and $\xi^{\alpha}\left(\bm{k}\right)$ are the $\alpha$-th eigenvalue (relative to the chemical potential $\mu_c$) and eigenvector of the TB model, respectively, and $n_F$ is the Fermi-Dirac distribution function.

\begin{figure}[htbp]
\centering
\includegraphics[width=0.4\textwidth]{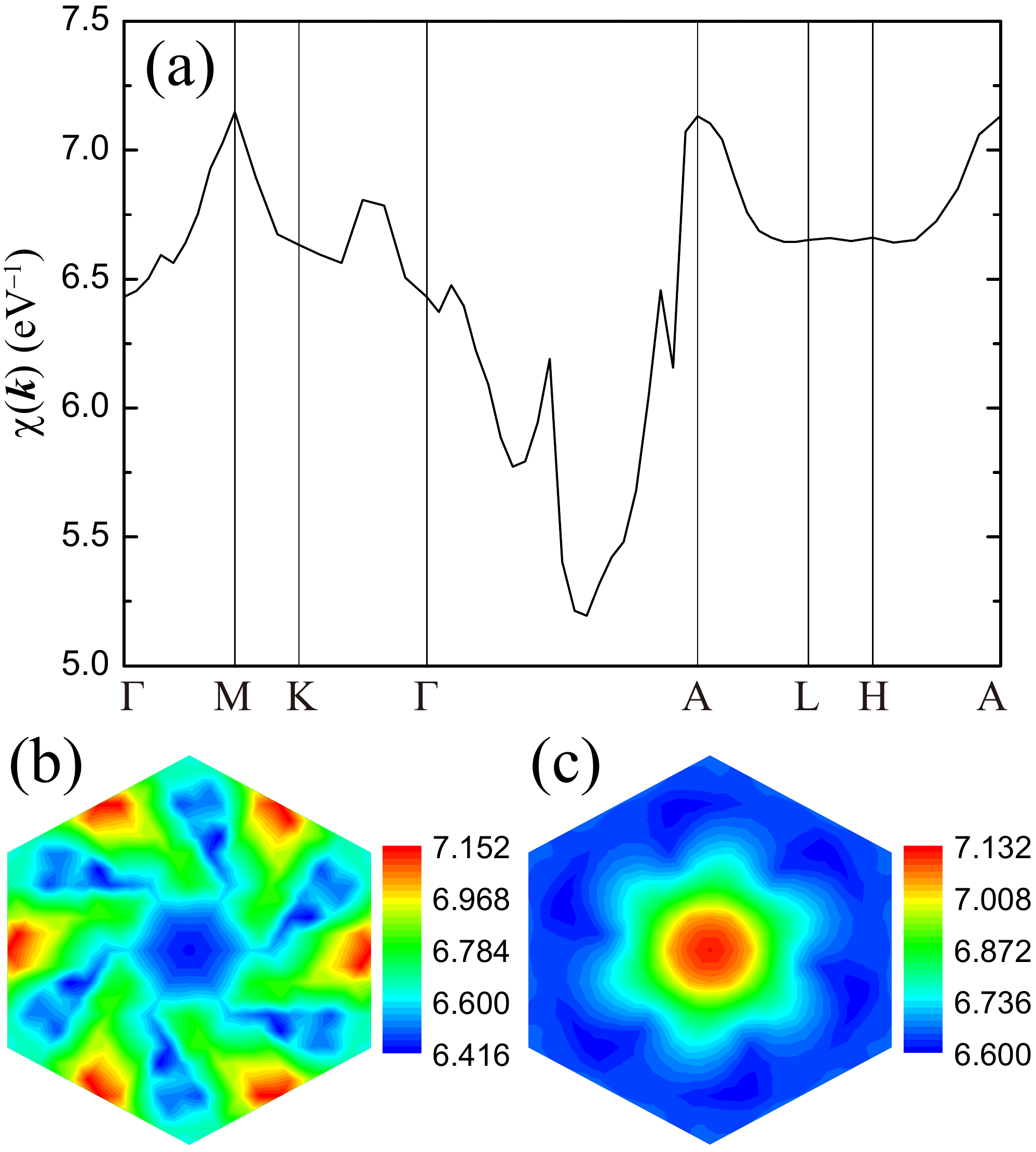}
\caption{(color online). The $\bm{k}$-space distribution of the largest eigenvalue of the susceptibility matrix $\chi^{(0)pq}_{st}(\bm{k},i\omega_n=0)$ (a) along the high-symmetry lines in the Brillouin zone, (b) on the $k_z=0$ plane and (c) on the $k_z=\pi$ plane.}
\label{chi}
\end{figure}

In Fig. \ref{chi}, we show the $\bm{k}$-dependence of the largest eigenvalue of the susceptibility matrix $\chi^{(0)}_{pqst}(\bm{k},i\omega_n=0)$ along the high-symmetry lines in the Brillouin zone in (a), on the $k_z=0$ plane in (b) and on the $k_z=\pi$ plane in (c). Clearly, the largest eigenvalue peaks near the A-point $(0,0,\pi)$ and the  M-points $(0,\pi,0), (\pi,0,0), (\pi,\pi,0)$. Detailed inspection reveals that the susceptibility peak near the A-point originates mainly from the nesting between the top and bottom surfaces of the 1D $\gamma,\delta$ FSs while that near the M-points originates from the nesting between different side surfaces of the 3D $\alpha,\beta$ FSs. The nearly equal peak values between the A- and M- points suggest competing wave vectors at these momenta for the density wave orders emerging at sufficiently strong interactions. Besides, there are a few other peaks appearing in Fig.~\ref{chi}, including the one centering at $(0,0,0.83\pi)$ near the A-point, which originates mainly from the nesting between the top (bottom) surfaces of the 3D $\alpha,\beta$ FSs and the bottom (top) surfaces of the 1D $\gamma,\delta$ FSs. This peak is intimately related to the origin of the $s^{\pm}$-wave pairing obtained below.

We further define the spin $(s)$ and charge $(c)$ susceptibilities. At the RPA level, the renormalized spin and charge susceptibilities of the system read
\begin{align}\label{chisce}
\chi^{(s,c)}(\bm{k},i\omega_n)=[I\mp\chi^{(0)}(\bm{k},i\omega_n)
U^{(s,c)}]^{-1}\chi^{(0)}(\bm{k},i\omega_n),
\end{align}
Here the nonzero elements $U^{(s,c)\mu\nu}_{\theta\xi}$ of $U^{(s,c)}$ satisfy $\mu,\nu,\theta,\xi\leq 3$ or $>3$ simultaneously, which are as follow,
\begin{align}
U^{(s(c))\mu\nu}_{\theta\xi}=\left\{
\begin{array}{ll}
U(U),  & \mu=\nu=\theta=\xi; \\
J_H(2V-J_H), & \mu=\nu\neq\theta=\xi; \\
J_H(J_H), & \mu=\theta\neq\nu=\xi; \\
V(2J_H-V),  & \mu=\xi\neq\theta=\nu.
\end{array}
\right.
\end{align}
In Eq.(\ref{chisce}), $\chi^{(s,c,0)}(\bm{k},i\omega_n)$ and $U^{(s,c)}$ are operated as $6^{2}\times 6^{2}$ matrices (see for example in Ref\cite{Liu2013}).

Generally, repulsive $U^{(s,c)}$ enhances $\chi^{(s)}$ and suppresses $\chi^{(c)}$. Increasing $U$ to some critical value $U_c$ (fixing $J_H/U$), $\chi^{(s)}(\bm{k},i\omega_n=0)$ will diverge first at the momentum ${\bs k=\bs Q=(0,0,\pi)}$, in close competition with the momenta ${\bs k=\bs Q=(\pi,0,0),(0,\pi,0)}$ and ${\bs (\pi,\pi,0)}$, suggesting the onset of SDW order with the corresponding wavevectors. Note that the SDW wavevectors $\bs Q$ in the KCr$_3$As$_3$ obtained here suggest inter-unit-cell SDW order, which stands in strong contrast with the inter-unit-cell ferromagnetic order with $\bs Q=(0,0,0)$ obtained for the A$_2$Cr$_3$As$_3$ family\cite{Wu:15}. The RPA approach is only valid for $U<U_c$.

At $U<U_c$, Cooper pairing may develop through exchanging spin and/or charge fluctuations. In particular, we consider Cooper pair scatterings both within and between the bands, hence both intra- and inter-band effective interactions $V^{\alpha\beta}(\mathbf{k,k'})$\cite{Wu:15} (here $\alpha/\beta=1,\cdots,6$ are band indices) are accounted for.  From the effective interaction vertex $V^{\alpha\beta}(\mathbf{k,k'})$, we obtain the following linearized gap equation (\ref{gapeq}):
\begin{align}\label{gapeq}
-\frac{1}{(2\pi)^3}\sum_{\beta}\oint_{FS}
d^{2}\bm{k}'_{\Vert}\frac{V^{\alpha\beta}(\bm{k},\bm{k}')}
{v^{\beta}_{F}(\bm{k}')}\Delta_{\beta}(\bm{k}')=\lambda
\Delta_{\alpha}(\bm{k}).
\end{align}
Here the integration runs along the $\beta$- FS, $v^{\beta}_F(\bm{k}')$ is the corresponding Fermi velocity, and $\bm{k}'_\parallel$ is the component of $\bm{k}'$ along the FS. Superconducting pairing in various channels emerge as the eigenstates of the above gap equation. The leading pairing $\Delta_\alpha(\bm{k})$ is given by the eigenstate corresponding to the largest eigenvalue $\lambda$. The critical temperature $T_c$ is related to $\lambda$ through $T_c\propto e^{-1/\lambda}$.

The eigenvector(s) $\Delta_{\alpha}(\bm{k})$ for each eigenvalue $\lambda$ obtained from gap equation (\ref{gapeq}) as the basis function(s) forms an irreducible representation of the $C_{6h}$ point group. In the absence of SOC, ten possible pairing symmetries are possible candidates for the system, which include five singlet pairings and five triplet pairings, as listed in Table.\ref{Tab:one}. For each triplet pairing, there are three degenerate components, i.e., $\uparrow\downarrow+\downarrow\uparrow$, $\uparrow\uparrow$, $\downarrow\downarrow$, and the degeneracy between them can be lifted by finite SOC.

\subsection{The phase diagram}
\label{sec:phase_diagram}

\begin{table}
\centering
\caption{The ten possible pairing symmetries for KCr$_3$As$_3$ in the absence of SOC, among which five are spin-singlet while the rest are spin-triplet.}
\label{Tab:one}
\begin{tabular}{@{}ccccccccccc@{}}
\\\hline\hline
 singlet   &&&&&&&&&&  triplet   \\
 \hline\hline
 $s$          &&&&&&&&&&  $p_z$  \\
 $(d_{x^2-y^2},d_{xy})$             &&&&&&&&&&      $(d_{x^2-y^2},d_{xy})\cdot p_z$             \\
 $(p_x,p_y)\cdot p_z$         &&&&&&&&&&  $(p_x,p_y)$          \\
 $f_{x^3-3xy^2}\cdot p_z$         &&&&&&&&&&  $f_{x^3-3xy^2}$             \\
 $f_{y^3-3x^2y}\cdot p_z$         &&&&&&&&&&  $f_{y^3-3x^2y}$             \\
 \hline\hline
\end{tabular}
\end{table}

\begin{figure}[htbp]
\centering
\includegraphics[width=0.5\textwidth]{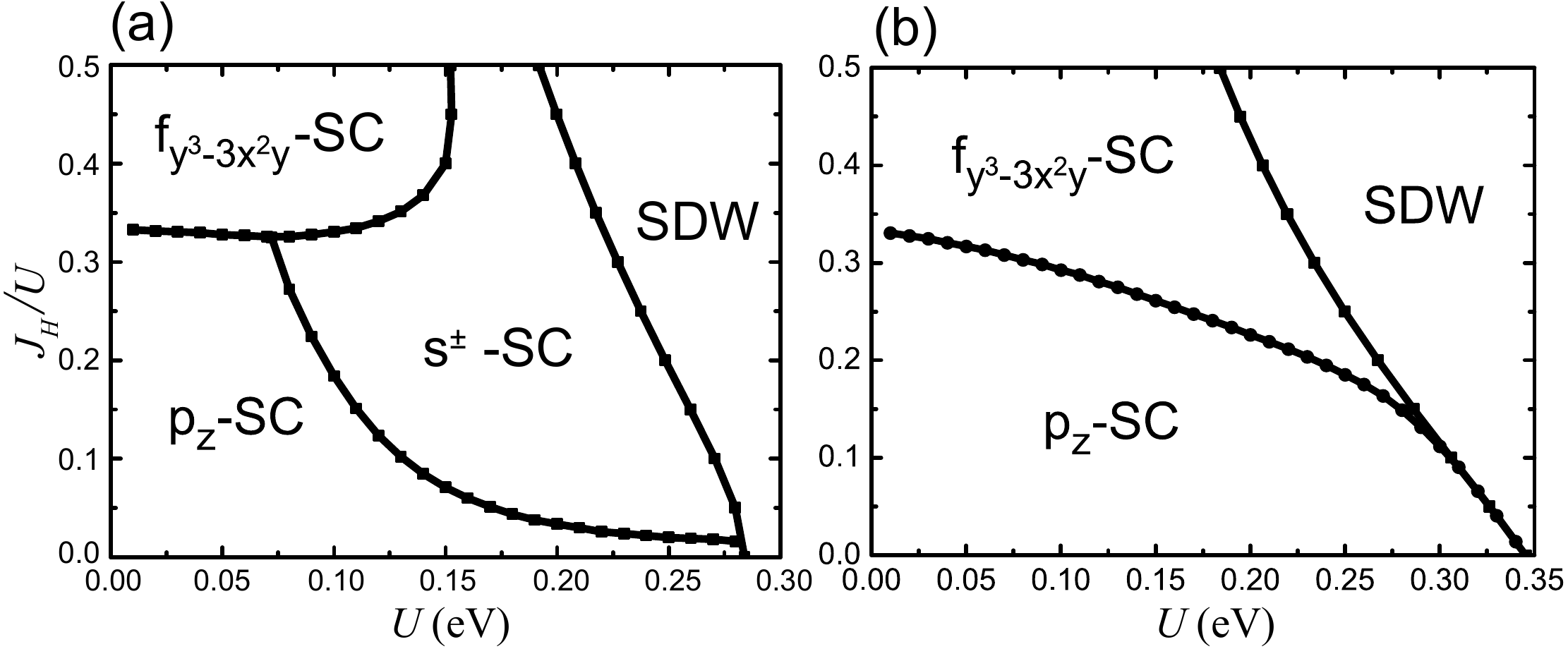}
\caption{(color online). The RPA phase diagram in the $U-J_H/U$ plane for (a) the KCr$_3$As$_3$ system studied here and (b) the K$_2$Cr$_3$As$_3$ system studied in Ref \onlinecite{Zhang:16}.}
\label{phase}
\end{figure}

\begin{figure}[htbp]
\centering
\includegraphics[width=0.48\textwidth]{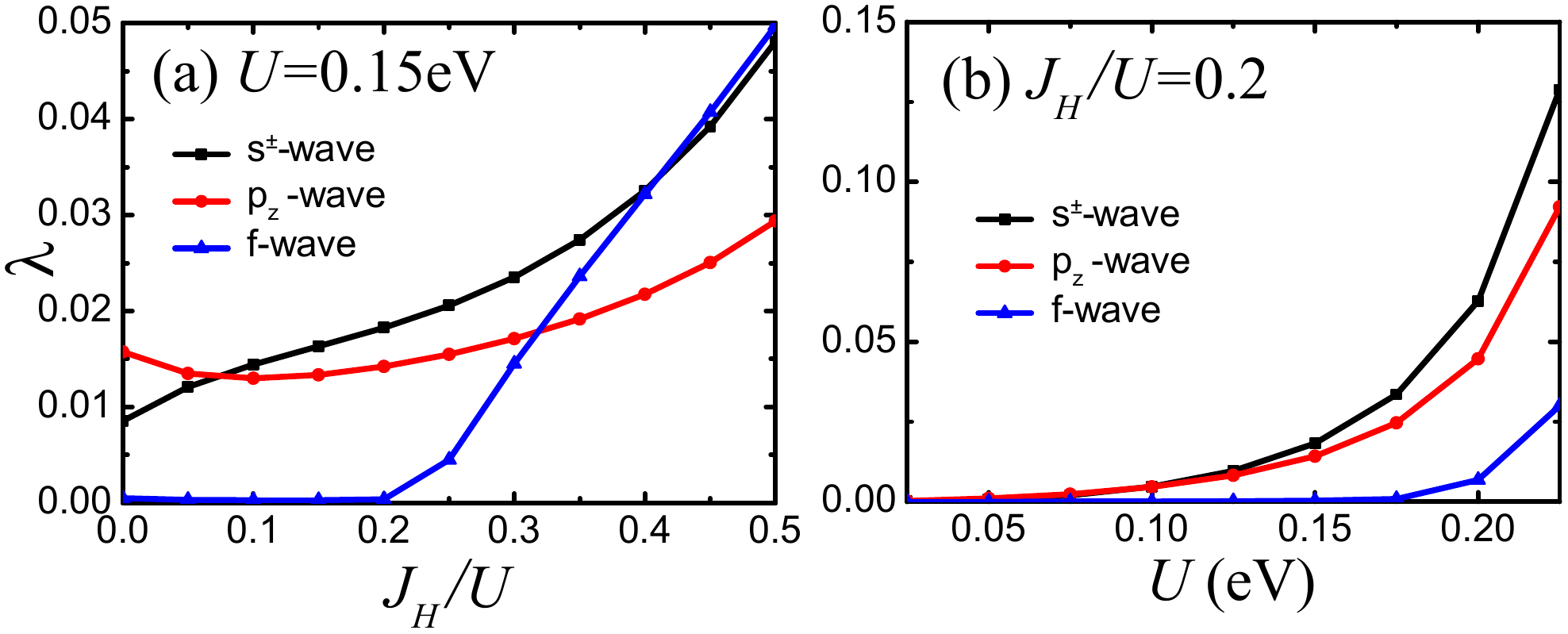}
\caption{(color online).(a) The $J_H/U$-dependence (fixing $U=0.15$eV) and (b) $U$-dependence (fixing $J_H/U=0.2$) of the largest eigenvalues $\lambda$ for all pairing symmetries of the system.}
\label{pair}
\end{figure}

\begin{figure}[htbp]
\centering
\includegraphics[width=0.5\textwidth]{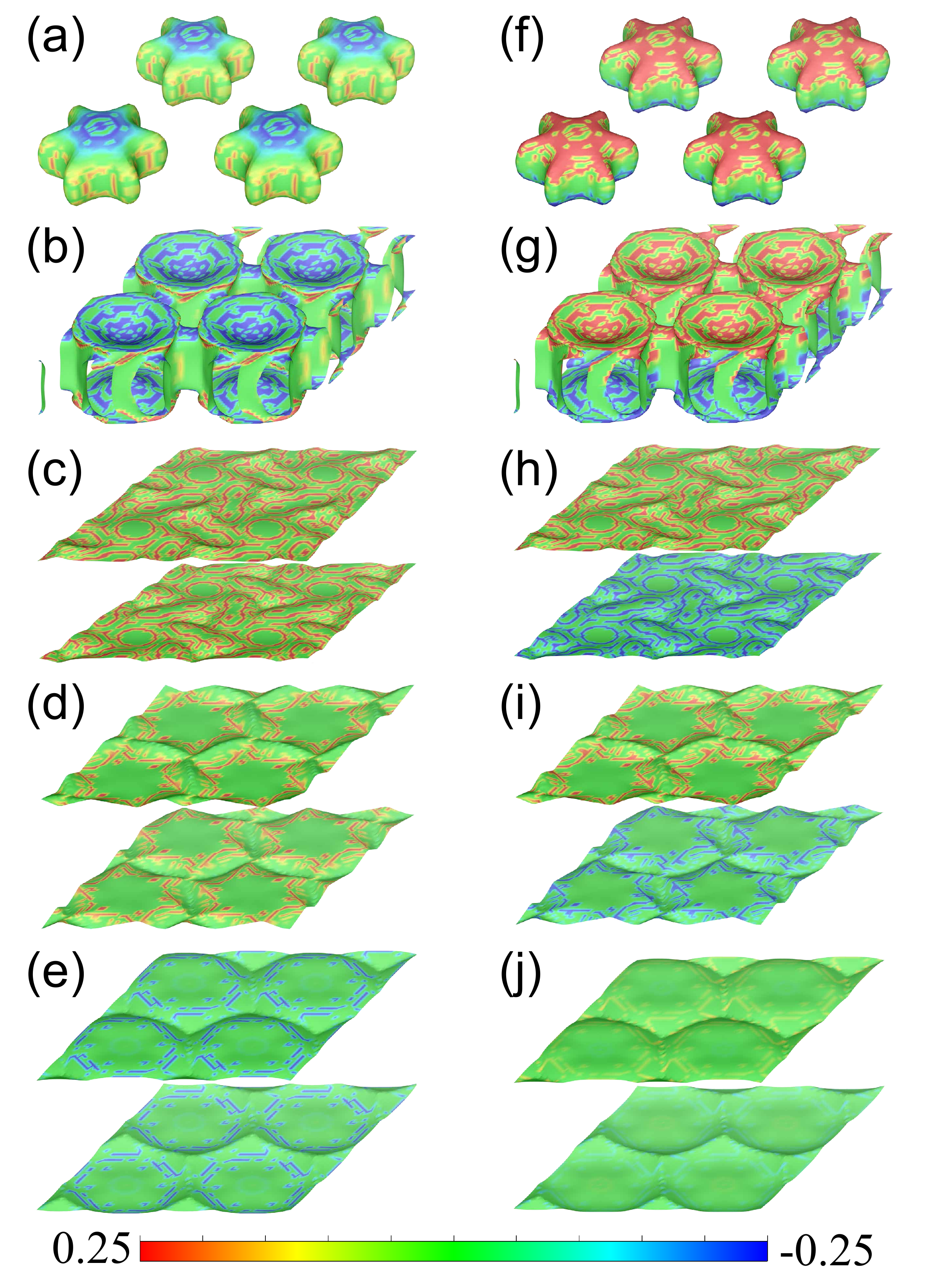}
\caption{(color online). The pairing gap functions shown on different FSs for the (a)-(e) $s^\pm$-SC with model parameters $U=0.2$eV, $J_H/U=0.1$, and (f)-(j) $p_z$-SC with parameters $U=0.05$eV, $J_H/U=0.1$. The FSs shown from the top to the bottom are  $\alpha,\beta,\gamma,\delta,\varepsilon$ respectively.}
\label{gap}
\end{figure}

The phase diagram as parameterized by the interactions $(U, J_H/U)$ is shown in Fig. \ref{phase}(a). Roughly speaking, SDW is favored at $U>U_c$, which is around $0.2\sim0.3$eV and is $J_H/U$-dependent in the present model. At weaker $U$, three superconducting phases appear, including spin-triplet $p_z$- and $f$-wave pairings and an $s$-wave pairing. The evolution of the different pairing channels are shown in Fig. \ref{pair}(a) as a function of $J_H/U$ at a representative interaction strength $U=0.15$eV, and in Fig. \ref{pair}(b) as a function of $U$ at a representative $J_H/U=0.2$. From the above results one finds: at weak $U$ the $f$-wave pairing becomes more favorable at sufficiently strong Hund's rule coupling ($J_H/U>r_c\approx \frac{1}{3}$), while $p_z$-wave pairing develops for weaker Hund's rule coupling; at intermediate $U$, the $s$-wave pairing dominates. For comparison, Fig. \ref{phase}(b) shows the phase diagram obtained from a previous study of the 233-family\cite{Zhang:16}.

The most striking difference between Fig. \ref{phase}(a) and (b) is the occurrence of $s$-wave pairing in the 133-family at intermediate values of $U$ and $J_H/U$. The normalized gap functions of the $s$-wave pairing on the five FSs are shown in Fig.\ref{gap}(a)-(e). The gap function is invariant under all $C_{6h}$ symmetry operations, and the gap amplitudes on all the five FSs are comparable. While the gap is positive on the two quasi-1D $\gamma$- and $\delta$-bands, it is negative on the quasi-1D $\varepsilon$-band. On the 3D $\alpha$- and $\beta$-bands, the gap function acquires opposite signs on the top/bottom and the side surfaces, giving rise to accidental nodal lines. The sign-changing character resembles the $s^\pm$-pairing proposed for iron-based SCs. In the present work, the $s^\pm$-pairing is driven by the SDW fluctuations and its sign structure is roughly consistent with the FS-nesting behavior as we discussed in association with Fig.\ref{chi}. To be more concrete, the nesting vector near ${\bs Q=(0,0,0.83\pi)}$ connects the top/bottom surfaces of the 3D $\alpha,\beta$ FSs and the bottom/top surfaces of the 1D $\gamma,\delta$ bands. As a result, the gap function on the corresponding nested portions of the FSs develops opposite signs.

The $p_z$-wave pairing for weak $U$ and $J_H/U$ shown in Fig. \ref{phase}(a) in the 133 family is similar to that in the 233 family\cite{Wu:15,Zhang:16} shown in Fig. \ref{phase}(b). Although the Cooper pairing in the weak $U$ and $J_H/U$ limit is still driven by spin fluctuations, the pairing symmetry obtained can be markedly different from that at intermediate $U$, because in this limit the bare interaction dominates the effective interaction. Although the dominant bare interaction in the weak $U$ limit will suppress any pairing symmetry, the $s$-wave will be most suppressed due to the following reason. The gap function of an $s$-wave (either $s^{++}$ or $s^\pm$) pairing have a nonzero average value over the entire Brilloui zone. As a result, when Fourier-transformed back into the real space, the $s$-wave pairing will inevitably possess a nonzero on-site pairing component, which is unfavored by the dominant repulsive on-site bare interaction in the weak $U$ limit with $J_H/U<r_c$. Hence, in the weak-$U$ limit with $J_H/U<r_c$, non-s-wave pairing is generally more favorable, and in the present study, the $p_z$-wave emerges as the leading pairing instability. Fig. \ref{gap}(f)-(j) show the normalized gap function in this channel. This pairing is invariant under a $C_6$ rotation, but changes sign under a mirror reflection about the $xy$ plane. As a result, line nodes exist in the $xy$ plane on the 3D bands. Note that the pairing amplitude is much weaker on the $\varepsilon$-band than the other bands, presumably due to a relatively weak interband interaction involving the $\varepsilon$-band.

For $J_H/U>r_c$, the $f$-wave pairing in the 133 and 233-families also share the same origin: the orbitals become unstable to Cooper pairing at the bare level of the on-site Hubbard interactions. This pairing possesses a significant on-site pairing component. This pairing is even about the $k_z = 0$ mirror plane, but changes sign under $C_{6}$ rotations. Therefore it also possesses line nodes.

{The three degenerate components of the triplet $p_z$- and $f$-wave pairings are split once spin-rotational invariance is broken by SOC. In the present system, to leading order the symmetry-allowed SOC takes the following form,
\begin{align}\label{chi0}
H_{SOC}=i\lambda_{SOC}&\sum_{\sigma, i\in A,B}
\sigma(c^{\dagger}_{i2\sigma}c_{i3\sigma}
-c^{\dagger}_{i3\sigma}c_{i2\sigma})
\end{align}
where 2, 3 stand for the orbital indices, and we take $\lambda_{SOC}=10$meV. Classified according to the $S_z$ quantum number of Cooper pairing, possible pairing components include the $S_z=0$, i.e., $\uparrow\downarrow+\downarrow\uparrow$ component, and the $S_z=\pm 1$, i.e., $\uparrow\uparrow$ and $\uparrow\uparrow$ component. The $J_H/U$ dependence of the largest pairing eigenvalues $\lambda$ of the $S_z=0$ and $S_z=\pm 1$ components of the $p_z$- and the $f$-wave pairings obtained from RPA is shown in Fig. \ref{soc} with fixed $U=0.05$eV. Clearly, for the $p_z$-wave, the $S_z=0$ component wins over the $S_z=\pm 1$ one for any $J_H/U$. As for the $f$-wave, the $S_z=0$ component wins over the $S_z=\pm 1$ one only for small $J_H/U$, and the contrary is true for large $J_H/U$.

\begin{figure}[htbp]
\centering
\includegraphics[width=0.4\textwidth]{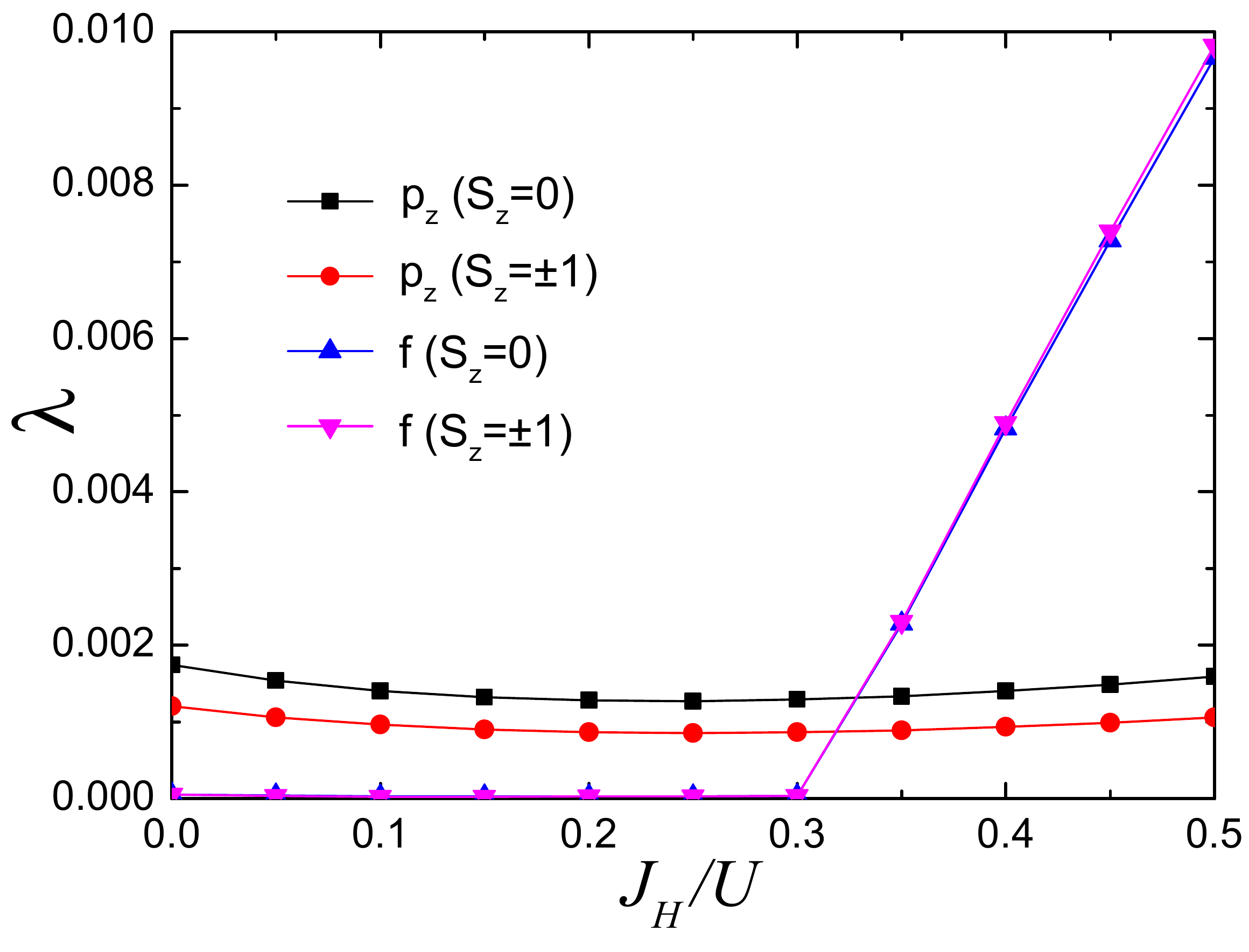}
\caption{(color online). The $J_H/U$-dependence of the largest eigenvalues $\lambda$ of the $S_z=0$ and $S_z=\pm 1$ components of the $p_z$- and $f$-wave pairings for $U=0.05$eV.}\label{soc}
\end{figure}

\section{The $s^\pm$-pairing}
\label{sec:sWave}
Here we argue that among the various phases exhibited in Fig. \ref{phase}(a), the most possible phase realized in KCr$_3$As$_3$ is the singlet $s^\pm$-wave SC. Firstly, the material is a superconductor without magnetic order, which excludes the SDW phase. Secondly, the realization of the $f$-wave SC requires large $J_H/U>r_c\approx\frac{1}{3}$, which is unrealistic. Thirdly, at weak $U$ where the $p_z$-wave emerges, the corresponding `expected' superconducting critical temperature $T_c$ is extremely low. As shown in Fig. \ref{pair} (b), fixing $J_H/U$ to a representative value 0.2, one sees that the $p_z$-wave SC becomes the leading pairing instability only  for $U<0.1$, which amounts to $\lambda<0.01$ and $T_c < e^{-100}$K. The result for other values of $J_H/U$ is similar. Such a low $T_c$ apparently contradicts with the experiments. Therefore, the $s^\pm$-wave SC driven by SDW fluctuations is left as the most probable candidate for KCr$_3$As$_3$.

Note that the $s^\pm$-SC obtained here by RPA has accidental line nodes on the two 3D bands $\alpha$ and $\beta$, as shown in Fig. \ref{gap}(a) and (b). As a result, some physical properties of this pairing cannot be easily distinguished from those of higher angular momentum, such as the $p_z$-wave pairing proposed in K$_2$Cr$_3$As$_3$. For example, in a pure sample, the differential conductance $dI/dV$ measured by the scanning-tunneling-microscopic (STM) should exhibit a $V$-like shape instead of a $U$-like one; the specific-heat $C_v$, the superfluid density $\rho_{s}$ and the NMR relaxation-rate $1/T_1T$ should scale in power-law with temperature $T$ at sufficiently low temperatures, instead of an exponential activation. Nevertheless, the spin-singlet $s^\pm$-wave state in KCr$_3$As$_3$ and the spin-triplet $p_z$-wave in K$_2$Cr$_3$As$_3$ shall be distinguishable in NMR Knight shift measurements \cite{Wu:1507}.

\begin{figure}[tb]
\centerline{\includegraphics[height=7 cm]{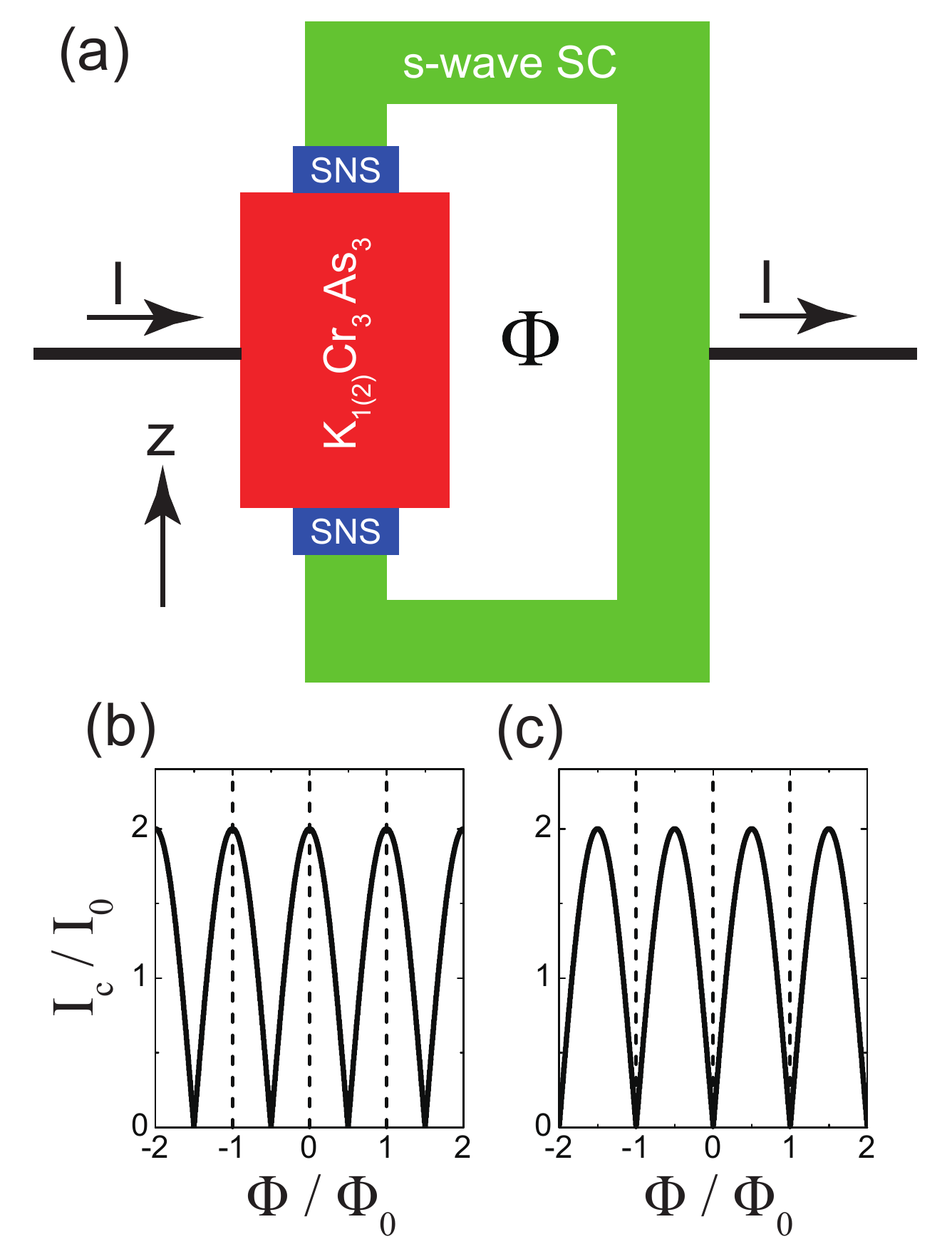}}
\caption{(color online)Sketched geometry for a SQUID phase sensitive probe (a) and the interference patterns of the SQUID for (b) KCr$_3$As$_3$ with $s$-wave pairing and (c) K$_2$Cr$_3$As$_3$ with $p_z$-wave pairing. \label{josephsonjunction} }
\end{figure}

\subsection{The phase-sensitive experiment}
\label{subsec:phase-sensitive}
The $s$-wave SC predicted here in the KCr$_3$As$_3$ can be distinguished from the $p_z$-wave predicted in the K$_2$Cr$_3$As$_3$ by the following dc SQUID device as shown in Fig.\ref{josephsonjunction}(a), where a loop of a conventional s-wave superconductor is coupled to the two ends of superconducting KCr$_3$As$_3$ or K$_2$Cr$_3$As$_3$ along the $z$ direction through superconductor-normal metal-superconductor (SNS) Josephson junctions. An externally applied magnetic flux $\Phi$ threads this loop. Due to the interference between the two branches of Josephson supercurrent, the maximum supercurrent (the critical current) $I_c$ in the circuit modulates with $\Phi$ according to
\begin{align}
I_c(\Phi)=2I_{0}\left|\cos\left(\pi\frac{\Phi}{\Phi_0}
+\frac{\delta\phi}{2}\right)\right|.
\end{align}
Here $I_0$ is the critical current of one Josephson junction, $\Phi_{0}=h/2e$ is the flux quantum. The phase shift $\delta\phi$ depends on the pairing symmetry. It is zero for an s-wave pairing and $\pi$ for $p_z$-wave. The different patterns of $I_c-\Phi$ relation can then be used to distinguish the two superconducting states, as demonstrated in Fig. \ref{josephsonjunction}.

\subsection{Spin resonance}
\label{subsec:spin_resonance}

\begin{figure}[htbp]
\centering
\includegraphics[width=0.45\textwidth]{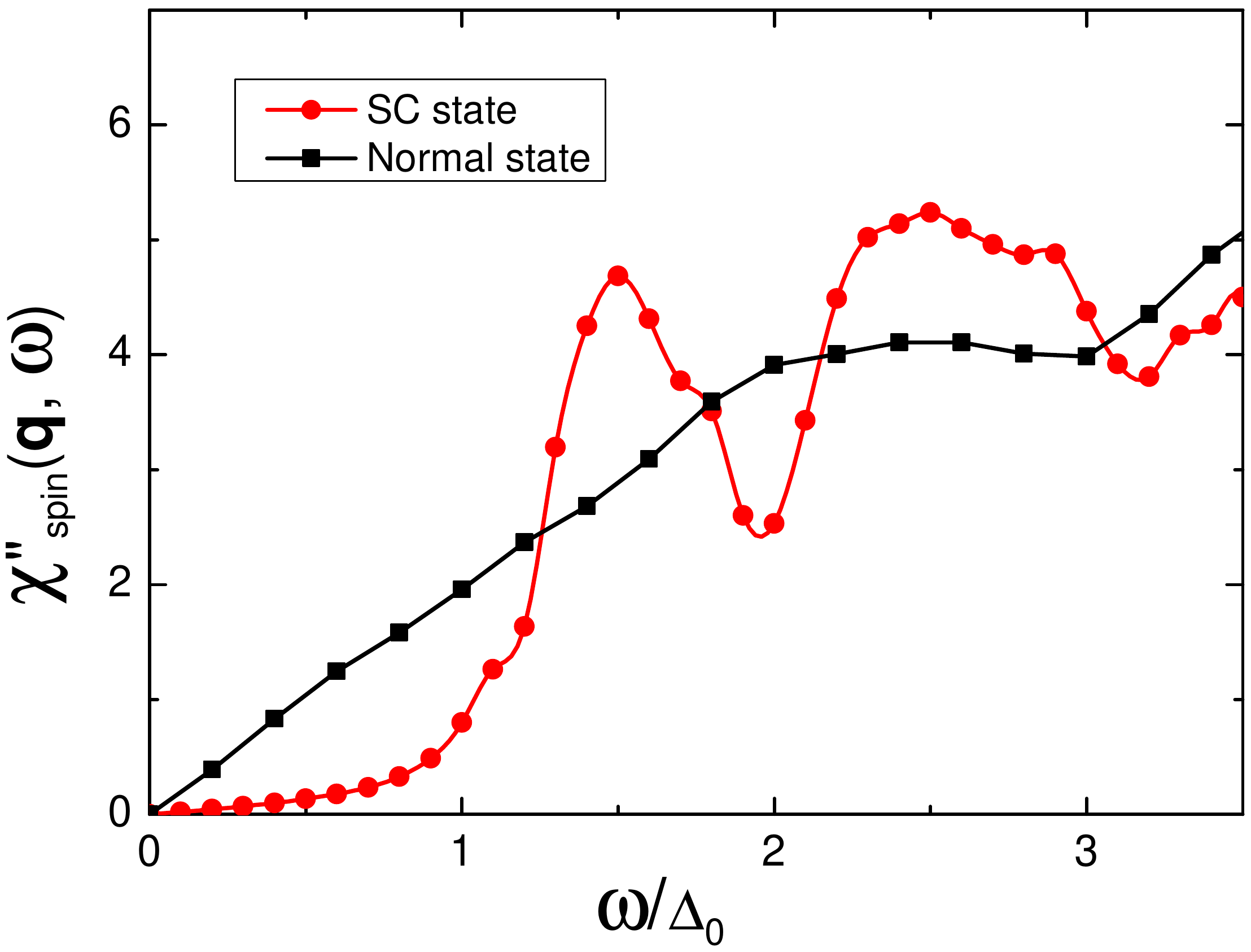}
\caption{(color online).The imaginary part of the dynamic spin susceptibility $\chi^{\prime\prime}_{\text{spin}}(\bm{q},\omega)\equiv\text{Im}\sum_{ps}\chi^{+-}_{ppss}(\bm{q},\omega)$ as a function of $\omega$ at a representative wavevector $\bm{q}=(0,0,0.83\pi)$ (which is near the nesting vector), for the normal state (black) and the $s^{\pm}$-wave state obtained in our RPA calculations (red). The frequence $\omega$ is measured in unit of $\Delta_0=2$meV. In these calculations we take $U=0.25$eV, $J_H/U=0.1$.}\label{spin_resonance}
\end{figure}
The above Josephson interferometry is oblivious to the sign-changing character of the $s$-wave pairing. However, there exists another distinguishing feature of the $s^\pm$-wave pairing -- the in-gap spin resonance mode in the superconducting state. The corresponding resonance peak emerges in the imaginary part of the dynamic spin susceptibility. Such a quantity can be easily obtained in the present RPA formalism.

Let's define the following bare susceptibility in the superconducting state,
\begin{eqnarray}
\chi^{(sc,0)ij}_{pqst}(\bm{q},\tau)&&\equiv
\frac{1}{4N}\sum_{\bm{k}_1\bm{k}_2}\left\langle
T_{\tau}c_{p\alpha}^{\dagger}(\bm{k}_1,\tau)\sigma^{i}_{\alpha\beta}
c_{q\beta}(\bm{k}_1+\bm{q},\tau)\right.                      \nonumber\\
&&\left.\cdot c_{s\alpha^{\prime}}^{\dagger}(\bm{k}_2+\bm{q},0)\sigma^{j}_{\alpha^{\prime}\beta^{\prime}}
c_{t\beta^{\prime}}(\bm{k}_2,0)\right\rangle_{sc,0},\label{chi0_sc_0}
\end{eqnarray}
where $\langle\cdots\rangle_{sc,0}$ denotes the thermal average in the BCS mean-field state. In $\sigma^{i}_{\alpha\beta}$, $\alpha,\beta$ denote spin indices and $\sigma^{i}$ represents the $i$-th component of the Pauli matrix. In the spin-singlet pairing state, the spin SU(2) symmetry requires $\chi^{(sc,0)ij}_{pqst}(\bm{q},\tau)=\delta_{ij}\chi^{(sc,0)ii}_{pqst}(\bm{q},\tau)$, which allows us to only calculate $\chi^{(sc,0)+-}_{pqst}(\bm{q},\tau)$ and obtain $\chi^{(sc,0)zz}_{pqst}(\bm{q},\tau)=\frac{1}{2}\chi^{(sc,0)+-}_{pqst}(\bm{q},\tau)$.

The above bare susceptibility function $\chi^{(sc,0)+-}_{pqst}(\bm{q},\tau)$ is first Fourier transformed to the imaginary frequency space and then analytically continued to the real frequency axis via $i\omega_n\to \omega+i0^{+}$, after which we obtain the following retarded bare susceptibility:
\begin{widetext}
\begin{eqnarray}\label{chi0_sc_1}
\chi^{(sc,0)+-}_{pqst}&&(\bm{q},\omega)=\frac{1}{4N}\sum_{\bm{k},\alpha\beta}\frac{1}{E^{\alpha}_{\bm{k}}E^{\beta}_{\bm{k+q}}}\biggl\{\nonumber\\
&&\frac{n_{F}(-E^{\alpha}_{\bm{k}})-n_{F}(-E^{\beta}_{\bm{k+q}})}{\omega+E^{\alpha}_{\bm{k}}-E^{\beta}_{\bm{k+q}}+i0^{+}}\cdot\left[(E^{\alpha}_{\bm{k}}+
\varepsilon^{\alpha}_{\bm{k}})(E^{\beta}_{\bm{k+q}}+\varepsilon^{\beta}_{\bm{k+q}})M^{(1)\alpha\beta}_{pqst}(\bm{k,q})+\Delta^{\alpha}_{\bm{k}}
\Delta^{\beta}_{\bm{k+q}}M^{(2)\alpha\beta}_{pqst}(\bm{k,q})\right]\nonumber\\&+&\frac{n_{F}(-E^{\alpha}_{\bm{k}})-n_{F}(E^{\beta}_{\bm{k+q}})}
{\omega+E^{\alpha}_{\bm{k}}+E^{\beta}_{\bm{k+q}}+i0^{+}}\cdot\left[(E^{\alpha}_{\bm{k}}+
\varepsilon^{\alpha}_{\bm{k}})(E^{\beta}_{\bm{k+q}}-\varepsilon^{\beta}_{\bm{k+q}})M^{(1)\alpha\beta}_{pqst}(\bm{k,q})-\Delta^{\alpha}_{\bm{k}}
\Delta^{\beta}_{\bm{k+q}}M^{(2)\alpha\beta}_{pqst}(\bm{k,q})\right]\nonumber\\&+&\frac{n_{F}(E^{\alpha}_{\bm{k}})-n_{F}(-E^{\beta}_{\bm{k+q}})}
{\omega-E^{\alpha}_{\bm{k}}-E^{\beta}_{\bm{k+q}}+i0^{+}}\cdot\left[(E^{\alpha}_{\bm{k}}-
\varepsilon^{\alpha}_{\bm{k}})(E^{\beta}_{\bm{k+q}}+\varepsilon^{\beta}_{\bm{k+q}})M^{(1)\alpha\beta}_{pqst}(\bm{k,q})-\Delta^{\alpha}_{\bm{k}}
\Delta^{\beta}_{\bm{k+q}}M^{(2)\alpha\beta}_{pqst}(\bm{k,q})\right]\nonumber\\&+&\frac{n_{F}(E^{\alpha}_{\bm{k}})-n_{F}(E^{\beta}_{\bm{k+q}})}
{\omega-E^{\alpha}_{\bm{k}}+E^{\beta}_{\bm{k+q}}+i0^{+}}\cdot\left[(E^{\alpha}_{\bm{k}}-
\varepsilon^{\alpha}_{\bm{k}})(E^{\beta}_{\bm{k+q}}-\varepsilon^{\beta}_{\bm{k+q}})M^{(1)\alpha\beta}_{pqst}(\bm{k,q})+\Delta^{\alpha}_{\bm{k}}
\Delta^{\beta}_{\bm{k+q}}M^{(2)\alpha\beta}_{pqst}(\bm{k,q})\right]\biggr\},
\end{eqnarray}
\end{widetext}
with
\begin{eqnarray}\label{M12}
E^{\alpha}_{\bm{k}}&=&\sqrt{(\varepsilon^{\alpha}_{\bm{k}})^2+\left|\Delta^{\alpha}_{\bm{k}}\right|^2},
\end{eqnarray}
and
\begin{eqnarray}\label{M12}
M^{(1)\alpha\beta}_{pqst}(\bm{k,q})&=&\xi^{\alpha*}_{p}(\bm{k})\xi^{\alpha}_{t}(\bm{k})\xi^{\beta}_{q}(\bm{k}+\bm{q})\xi^{\beta*}_{s}(\bm{k}+\bm{q}),\nonumber\\
M^{(2)\alpha\beta}_{pqst}(\bm{k,q})&=&\xi^{\alpha*}_{p}(\bm{k})\xi^{\alpha}_{s}(\bm{k})\xi^{\beta}_{q}(\bm{k}+\bm{q})\xi^{\beta*}_{t}(\bm{k}+\bm{q}).
\end{eqnarray}
Here the pairing gap $\Delta^{\alpha}_{\bm{k}}$ is a product of the dimensionless normalized gap function $\Delta_{\alpha}(\bm{k})$ solved in Eq.(\ref{gapeq}) and a global amplitude $\Delta_0$, i.e. $\Delta^{\alpha}_{\bm{k}}=\Delta_0\Delta_{\alpha}(\bm{k})$ ($|\Delta_{\alpha}(\bm{k})|_{\text{max}}=1$). Here, we take $\Delta_0=2$meV on account of the experimental superconducting $T_c\approx 5$ K.

The RPA susceptibility in the superconducting state is still defined by Eq.(\ref{chisce}), but with $\chi^{(0)}(\bm{k},i\omega_n)$ replaced by $\chi^{(sc,0)+-}(\bm{k},\omega)$ in Eq.(\ref{chi0_sc_1}). The quantity related to the inelastic neutron scattering experiment is the imaginary part $\chi^{\prime\prime}_{\text{spin}}(\bm{q},\omega)$ of the dynamic spin susceptibility $\chi^{+-}_{\text{spin}}(\bm{q},\omega)\equiv\sum_{ps}\chi^{+-}_{ppss}(\bm{q},\omega)$. The obtained $\chi^{\prime\prime}_{\text{spin}}(\bm{q},\omega)\equiv \text{Im}\chi^{+-}_{\text{spin}}(\bm{q},\omega)$ at zero temperature is shown in Fig. \ref{spin_resonance} as a function of $\omega$ for a typical fixed momentum $\bm{q}=(0,0,0.83\pi)$ near the nesting vector, in comparison with the renormalized results obtained in the non-superconducting normal state. Fig.~\ref{spin_resonance} shows a pronounced spin resonance peak below $w=2\Delta_0$ in the $s^{\pm}$-wave state, which is absent in the normal state. Physically, the presence of this collective spin mode not only confirms the sign-changing character of the $s^{\pm}$-wave pairing in the 133 family, but also indicates the magnetic origin of the Cooper pairing. This resonance mode can be easily detected in inelastic neutron scattering measurements.

\subsection{The TRI TSC}
\label{subsec:TRI_TSC}

\begin{figure}[htbp]
\centering
\includegraphics[width=0.4\textwidth]{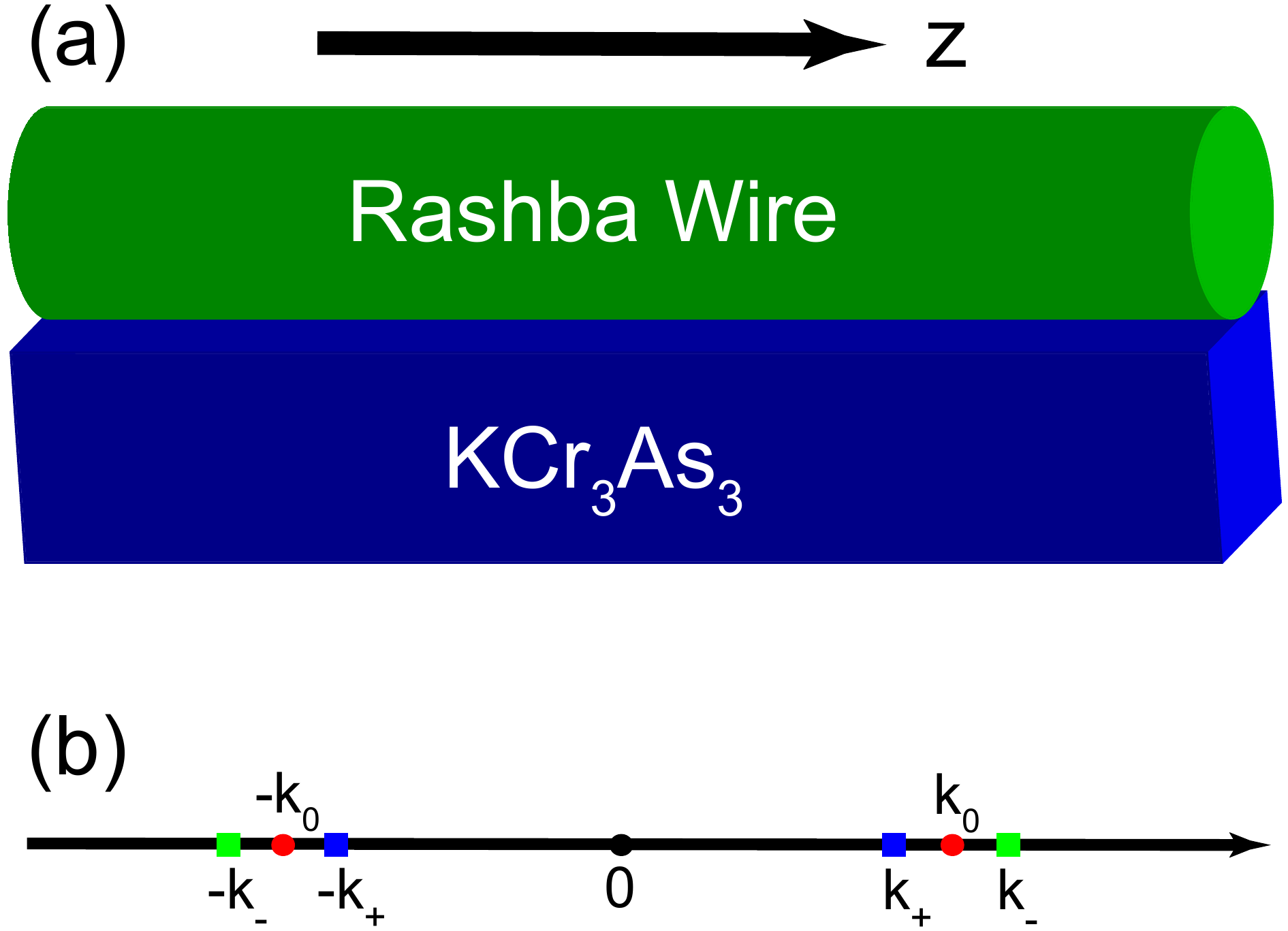}
\caption{(color on line). (a) Sketch of the proximity device between the KCr$_3$As$_3$ (blue) and a Rashba semiconductor wire(green). (b) Sketch of nagative pairing.}
\label{topological_SC}
\end{figure}

A remarkable property of the quasi-1D $s^\pm$-SC obtained here lies in that, similar with the case of quasi-2D $s^\pm$-wave iron-based SC\cite{FanZhang}, it can be equipped to engineer 1D TRI TSC via proximity effects between the $s^\pm$-wave KCr$_3$As$_3$ superconductor and a semiconductor wire with large Rashba SOC. As shown in Fig. \ref{topological_SC}(a), the quasi-1D KCr$_3$As$_3$ superconductor (blue) is parallel aligned close to a semiconductor wire along the Cr-chain direction (marked as the $z$-axis direction). The semiconductor wire has large Rashba SOC $\lambda_R$ and similar lattice constant with KCr$_3$As$_3$. By proximity effect, the superconducting pairing in the KCr$_3$As$_3$ will be delivered to the Rashba wire. Due to the repulsive Hubbard-interactions in our model, the on-site pairing component $\Delta_0$ of the $s^\pm$-wave SC should be weak and the inter-site-pairing component will be dominant. For the quasi-1D KCr$_3$As$_3$ superconductor, the dominant inter-site-pairing component should be the pairing $\Delta_1$ between nearest-neighbor (NN) unit-cell along the Cr-chain. We assume such pairing component structure will be delivered to the 1D Rashba wire via proximity effect.

The Hamiltonian for the Rashba wire with SC acquired via proximity to KCr$_3$As$_3$ reads,
\begin{eqnarray}
H&=&-t\sum_{i\sigma}c^{\dagger}_{i\sigma}c_{i+1,\sigma}+h.c.-i\lambda_R\sum_{i}c^{\dagger}_{i}\sigma^{y}(c_{i+1}-c_{i-1})\nonumber\\
&&+\Delta_0\sum_{i}(c^{\dagger}_{i\uparrow}c^{\dagger}_{i\downarrow}+h.c.)+\Delta_1\sum_{i}\big[(c^{\dagger}_{i\uparrow}c^{\dagger}_{i+1,\downarrow}-
c^{\dagger}_{i\downarrow}c^{\dagger}_{i+1,\uparrow})\nonumber\\&&+h.c.\big]-\mu_c\sum_{i\sigma}c^{\dagger}_{i\sigma}c_{i\sigma}\label{proximity}
\end{eqnarray}
Here $t$ is the NN hopping coefficient, $\lambda_R$ is the coefficient of the Rashba SOC, $\mu_c$ is the chemical potential, $\Delta_0$ and $\Delta_1$ are the on-site and NN- pairing amplitudes respectively. After Fourier-transformed to the momentum space, the Hamiltonian can be written as
\begin{eqnarray}
H&=&\sum_{k}\left[\psi^{\dagger}_{k}h_{k}\psi_{k}+\frac{1}{2}\left(\psi^{\dagger}_{k}\Delta_{k}\psi^{\dagger T}_{-k}+h.c.\right)\right],\label{proximity_k}
\end{eqnarray}
with $\psi_{k}=\left(c_{k\uparrow}, c_{k\downarrow}\right)^{T}$, $h(k)=\left(-2t\cos k-\mu_c\right)I+2\lambda_R\sin k\sigma^{y}$ and $\Delta_{k}=i\sigma^{y}\Delta\left(k\right)$. Here $\Delta\left(k\right)=\Delta_0+2\Delta_1\cos k$ is the gap function. This Hamiltonian is TRI.

From the criterion of TRI TSC in 1D\cite{Xiaoliang}, we should calculate the following $Z_2$ FS topological invariant (FSTI),
\begin{equation}
N_{1D}=\prod_{s}[\text{sgn}(\delta_s)],\label{FSTI}
\end{equation}
where the summation index $s$ represents any Fermi point between $0$ and $\pi$, and the real number $\delta_s$ is defined as\cite{Xiaoliang}
\begin{equation}
\delta_s\equiv\left\langle n_{s},k_{s}\left|\mathcal{T}\Delta^{+}_{k}\right|n_{s},k_{s}\right\rangle.\label{delta}
\end{equation}
Here $\left|n_{s},k_{s}\right\rangle$ is the eigenvector of $h_k$, with $k_{s}$ and $n_{s}$ denoting the momentum and band index for the Fermi-point $s$, $\mathcal{T}=i\sigma^y$ is the time-reversal matrix\cite{Xiaoliang}, and $\Delta^{\dagger}_{k}=-i\sigma^y\Delta\left(k\right)$. As a result, we have,
\begin{equation}
\delta_s=\Delta(k_s).\label{delta2}
\end{equation}

For each $k$, the two eigenvalues $\varepsilon^{\pm}(k)$ of $h_k$ are solved as
$\varepsilon^{\pm}(k)=-2t\cos k-\mu_c\pm 2\lambda_R \sin k$. For each $\mu_c$, we have two Fermi-points $k_{\pm}$ within $(0,\pi)$, satisfying $\varepsilon^{\pm}(k_{\pm})=0$ respectively, with $k_{+}<k_{-}$. Therefore, the FSTI defined in Eq.(\ref{FSTI}) is written as
\begin{equation}
N_{1D}=\text{sgn}[\Delta(k_{+})]\text{sgn}[\Delta(k_{-})],\label{sign}
\end{equation}
from which the criterion of TRI TSC is that the signs of $\Delta(k_{\pm})$ are different, known as the negative pairing. Noticing that the node of the gap function $\Delta(k)=0$ within $(0,\pi)$ solved as $k_0=\arccos(-\Delta_0/2\Delta_1)$ is unique, the situation of negative pairing requires $k_+<k_0<k_-$ as shown in Fig. \ref{topological_SC}(b). This condition leads to $\mu_c\in(\frac{t\Delta_0}{\Delta_1}-2\lambda_R\sqrt{1-\frac{\Delta^{2}_0}{4\Delta^{2}_1}},\frac{t\Delta_0}{\Delta_1}+2\lambda_R\sqrt{1-\frac{\Delta^{2}_0}{4\Delta^{2}_1})}$. Particularly, the width of the energy window of $\mu_c$ required by TRI TSC is $4\lambda_R\sqrt{1-\frac{\Delta^{2}_0}{4\Delta^{2}_1}}$, which is non-zero when $\left|\frac{\Delta_1}{\Delta_0}\right|>\frac{1}{2}$. Such condition can be easily satisfied.

\section{Discussion and Conclusion}
\label{sec:summary}

\begin{figure}[htbp]
\centering
\includegraphics[width=0.48\textwidth]{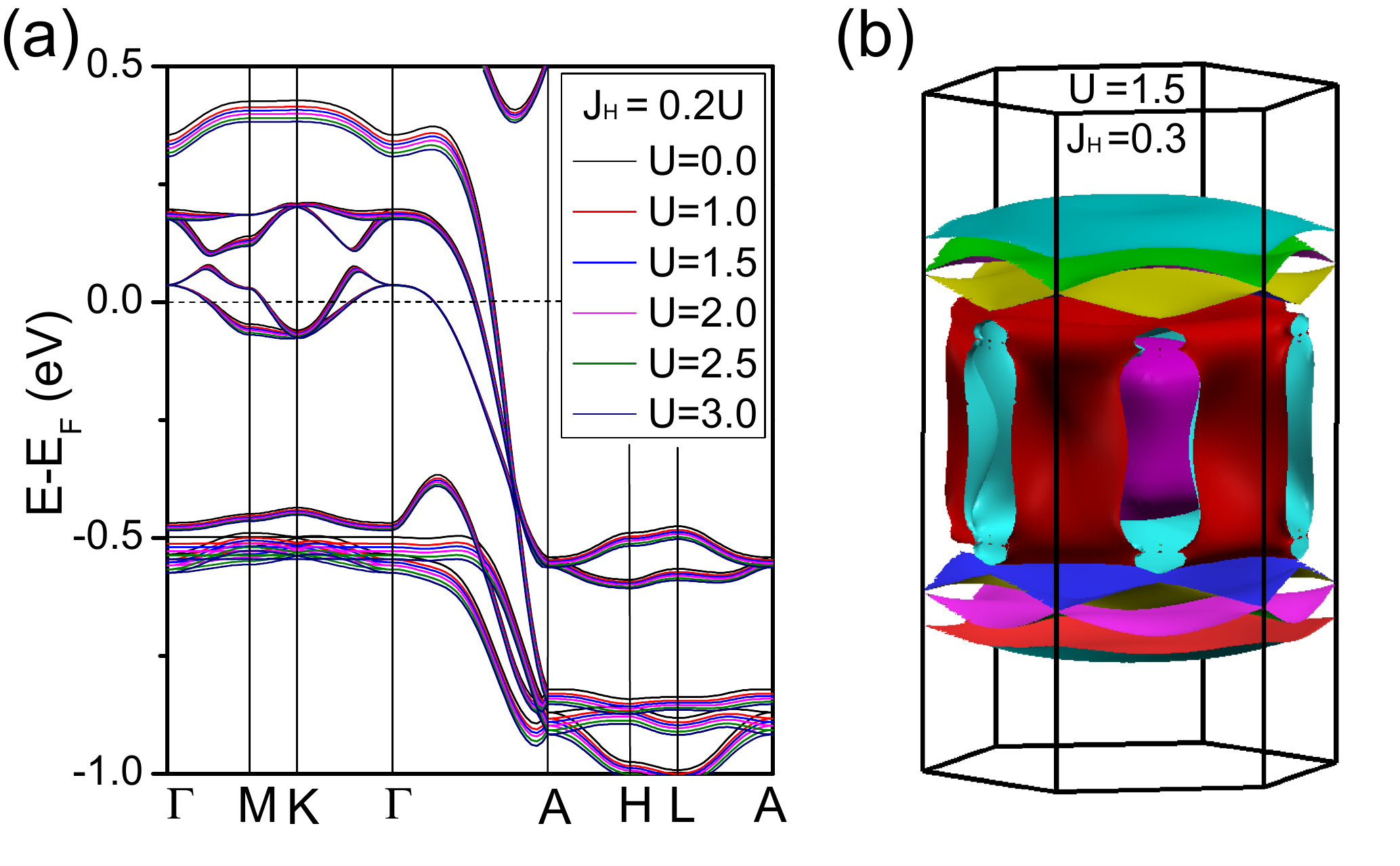}
\caption{(color online). The LDA+U band structures. (a) The $U$-dependent band structures, fixing $J_H=0.2U$. (b) The FS for fixed $U=1.5$eV, $J_H=0.3$eV. Note that the parameters $U$ and $J_H$ are defined on conventionally local ``atomic-orbital" Wannier bases, which are different from those appearing in Eq.(\ref{model})}
\label{LDA_plus_U}
\end{figure}

To check how the electron correlation affects the band structure of KCr$_3$As$_3$, we have performed an LDA+U calculation on the system. The on-site Coulomb interaction is added on the localized 3$d$ orbitals of Cr atoms (note that we adopt the "atomic-orbital" here) by using the method introduced by Dudarev $et$ $al.$ \cite{Dudarev} The results are shown in Fig.~\ref{LDA_plus_U}. Fixing $J_H=0.2U$, the band structures for $U=$0, 1.0eV, 1.5eV, 2.0eV, 2.5eV and 3.0eV are shown in Fig.~\ref{LDA_plus_U}(a), where one can easily verify that the low energy band structure near the Fermi level does not obviously change with $U$. The FS for fixed $U=1.5$eV, $J_H=0.3$eV is shown in Fig.~\ref{LDA_plus_U}(b), which is qualitatively the same as that for $U=0$ shown in Fig.~\ref{structure}(c), with slightly shortened nesting vector. All the conclusions attained here are maintained.

In conclusion, we have studied the pairing symmetry in the new Cr-based superconductor KCr$_3$As$_3$ based an effective model constructed from the input provided by DFT calculations. Our RPA analyses of the Hubbard-Hund model of this material reveals three possible pairing symmetries in different regimes of the interaction parameter space, i.e. the $f$-wave, the $p_z$-wave and the $s^\pm$-wave. For realistic parameters, we argue that the singlet $s^\pm$-wave pairing is the leading pairing symmetry. This singlet pairing is driven by SDW fluctuations enhanced by FS nesting. The pairing possesses accidental nodal lines on the 3D bands. We also pointed out phase sensitive measurement to identify the pairing symmetries in KCr$_3$As$_3$ and the related compound K$_2$Cr$_3$As$_3$. An interesting aspect of the $s^{\pm}$-SC is the presence of a spin resonance mode in the superconducting state. Signatures of this collective mode shall appear in inelastic neutron scattering experiment and can thus serve as a smokinggun evidence for the $s^\pm$-wave pairing predicted in the present work. Remarkably, this quasi-1D $s^\pm$-wave SC with dominant intra-chain inter-site pairing can be utilized to engineer 1D TRI TSC by proximately coupling it to a Rashba wire with large SOC. As a result of the nontrivial $Z_2$ FSTI, there should be a topologically protected Majorana Kramers pair on each end of the Rashba wire.

The present study mainly focuses on the weak-coupling regime. As for the strong-coupling case, the super-exchange interactions should be considered, instead of the Hubbard-Hund interactions studied here. Previous spin-dependent DFT calculations\cite{Cao133} suggest that the ground state of KCr$_3$As$_3$ is magnetically ordered with interlayer antiferromagnetic (AFM) structure. Although no magnetic order has been detected in experiments \cite{Mu:17, Liu:17}, the super-exchange interaction and the short-ranged AFM fluctuations should still be present in the material. As a result of the intra-chain inter-layer AFM super-exchange interactions, the quasi-1D $s$-wave pairing with dominant intra-chain inter-site pairing will be most favorable. We expect the pairing thus obtained to be similar to the $s^\pm$-wave pairing obtained in our RPA approach.

\section*{Acknowledgements}
We particularly thank K. M. Taddei for the suggestion of studying the spin resonance mode and Cheng-Cheng Liu for the suggestion of studying possible TRI TSC via proximity. We are also grateful to the stimulating discussions with Zhi-An Ren, Guo-Qing Zheng and Guang-Han Cao. This work is supported by the NSFC (Grant Nos. 11674025, 11604013, 11674151, 11334012, 11274041), Beijing Natural Science Foundation (Grant No. 1174019). Li-Da Zhang and Xiaoming Zhang contribute equally to this work.

\newpage
\widetext
\section{appendix}
The values of the hopping parameters appearing in the TB Hamiltonian in the main text are listed in the following.
\begin{table}[h]
\scriptsize
\caption{\label{hopping1} Note that $t^{0,n}_{33}=t^{0,n}_{22}$,
$\tilde{t}^{0,n}_{33}=\tilde{t}^{0,n}_{22}$.}
\begin{ruledtabular}
\begin{tabular}{ccccccccc}
$\mu\nu$&$t^{0,0}$&$t^{0,1}$&$t^{0,2}$&$t^{0,3}$&$\tilde{t}^{0,0}$&
$\tilde{t}^{0,1}$&$\tilde{t}^{0,2}$&$\tilde{t}^{0,3}$\\
\colrule
11 & 2.3956 & 0.1901 & -0.0761 & 0.0098 &  0.2148 & -0.0082 &  0.0064 &  0.0032 \\
22 & 2.5043 & 0.1619 & -0.0479 & 0.0143 & -0.0327 &  0.0350 & -0.0065 & -0.0033
\end{tabular}
\end{ruledtabular}
\end{table}

\begin{table}[h]
\scriptsize
\caption{\label{hopping2} }
\begin{ruledtabular}
\begin{tabular}{ccccccccc}
$\mu\nu$&$t^{1,0}$&$t^{1,1}$&$t^{1,2}$&$t^{1,3}$&
$t^{2,0}$&$t^{2,1}$&$t^{2,2}$\\
\colrule
11 &  0.0030 & -0.0028 & -0.0023 &  0.0010 & -0.0021 &  0.0014 &         \\
12 &         & -0.0021 & -0.0006 &  0.0013 &  0.0019 & -0.0007 &         \\
13 &  0.0033 & -0.0012 & -0.0006 &  0.0029 & -0.0006 & -0.0002 &         \\
21 & -0.0138 &  0.0188 &  0.0011 & -0.0016 &         &         &         \\
22 &  0.0017 &  0.0011 & -0.0024 &  0.0011 &  0.0002 &         &         \\
23 &  0.0035 &  0.0081 &  0.0022 & -0.0004 & -0.0020 &  0.0027 & -0.0010 \\
31 & -0.0045 &  0.0062 & -0.0020 &  0.0017 &         &         &         \\
32 & -0.0028 &  0.0018 &  0.0001 & -0.0030 &  0.0001 & -0.0002 &  0.0002 \\
33 &  0.0015 &  0.0031 &  0.0015 &  0.0003 &  0.0024 & -0.0021 &
\end{tabular}
\end{ruledtabular}
\end{table}

\begin{table}[h]
\scriptsize
\caption{\label{hopping3} }
\begin{ruledtabular}
\begin{tabular}{ccccccccc}
$\mu\nu$&$t^{1A,0}$&$t^{1A,1}$&$t^{1A,2}$&$t^{1A,3}$&
$t^{2A,0}$&$t^{2A,1}$&$t^{2A,2}$&$t^{2A,3}$\\
\colrule
11 &         & -0.0014 &         &         &         &         &         &         \\
12 & -0.0011 &  0.0052 & -0.0005 & -0.0003 & -0.0002 &  0.0006 &         &         \\
13 & -0.0045 &  0.0011 &  0.0010 &  0.0005 &  0.0003 & -0.0009 &         &         \\
21 & -0.0011 &  0.0052 & -0.0005 & -0.0003 & -0.0002 &  0.0006 &         &         \\
22 & -0.0170 & -0.0113 &  0.0003 &  0.0001 & -0.0005 &         & -0.0010 & -0.0005 \\
23 & -0.0043 & -0.0042 & -0.0001 & -0.0001 &  0.0004 &         &  0.0004 &  0.0002 \\
31 & -0.0045 &  0.0011 &  0.0010 &  0.0005 &  0.0003 & -0.0009 &         &         \\
32 & -0.0043 & -0.0042 & -0.0001 & -0.0001 &  0.0004 &         &  0.0004 &  0.0002 \\
33 & -0.0014 & -0.0034 &  0.0001 &         & -0.0004 &         & -0.0004 & -0.0002
\end{tabular}
\end{ruledtabular}
\end{table}

\begin{table}[h]
\scriptsize
\caption{\label{hopping4} }
\begin{ruledtabular}
\begin{tabular}{ccccccccc}
$\mu\nu$&$t^{1B,0}$&$t^{1B,1}$&$t^{1B,2}$&$t^{1B,3}$ &$t^{2B,1}$\\
\colrule
11 &  0.0033 & -0.0072 & 0.0025 & 0.0012 &         \\
12 & -0.0023 & -0.0015 & 0.0010 & 0.0005 & -0.0005 \\
13 &  0.0007 & -0.0071 & 0.0011 & 0.0005 &  0.0016 \\
21 & -0.0023 & -0.0015 & 0.0010 & 0.0005 & -0.0005 \\
22 &  0.0034 & -0.0055 & 0.0035 & 0.0017 & -0.0005 \\
23 &  0.0003 & -0.0020 & 0.0009 & 0.0004 & -0.0005 \\
31 &  0.0007 & -0.0071 & 0.0011 & 0.0005 &  0.0016 \\
32 &  0.0003 & -0.0020 & 0.0009 & 0.0004 & -0.0005 \\
33 &  0.0041 &         & 0.0035 & 0.0017 & -0.0012
\end{tabular}
\end{ruledtabular}
\end{table}

\end{document}